\documentclass[prb,twocolumn,preprintnumbers,amsmath,amssymb]{revtex4}

\usepackage{graphicx}
\usepackage{dcolumn}
\usepackage{bm}
\usepackage{mathrsfs}
\usepackage{bbm}
\usepackage{color}

\def\E{\rm E}

\def\S{\rm S}
\def\be{\begin{equation}}
\def\ee{\end{equation}}
\def\bd{\begin{displaymath}}
\def\ed{\end{displaymath}}
\def\-{\phantom{-}}

\begin{document}
%
%

\title{Spin-lozenge thermodynamics and magnetic excitations in Na$_3$RuO$_4$}

\author{J. T. Haraldsen$^{1,4}$, M. B. Stone$^{2}$, M. D. Lumsden$^{2}$, E. Mamontov$^{2}$, T. Barnes$^{1,3}$, R. Jin$^4$, J. W. Taylor$^5$, and F. Fernandez-Alonso$^5$}
\affiliation{$^1$Department of Physics and Astronomy, University of Tennessee, Knoxville Tennessee 37996, USA}
\affiliation{$^2$Neutron Scattering Science Division, Oak Ridge National Laboratory, Oak Ridge, Tennessee 37831, USA}
\affiliation{$^3$Physics Division, Oak Ridge National
Laboratory, Oak Ridge, Tennessee 37831, USA}
\affiliation{$^4$Materials Science and Technology Division, Oak Ridge National Laboratory, Oak Ridge, Tennessee 37831, USA}
\affiliation{$^5$ISIS Facility, Rutherford Appleton Laboratory, Chilton, Didcot, Oxfordshire OX11 0QX, United Kingdom}

\date{\today}

\begin{abstract}

We report inelastic and elastic neutron scattering, magnetic susceptibility, and heat
capacity measurements of polycrystalline sodium ruthenate (Na$_3$RuO$_4$).
Previous work suggests this material consists of isolated tetramers of $S=3/2$
Ru$^{5+}$ ions in a so-called ``lozenge" configuration.  Using a Heisenberg
 antiferromagnet Hamiltonian, we analytically determine the
energy eigenstates for general spin $S$. From this model, the neutron scattering cross-sections for excitations
associated with spin-3/2 spin-tetramer
configurations is determined. Comparison of magnetic susceptibility and inelastic neutron
scattering results shows that the proposed ``lozenge" model is not distinctly supported,
but provides evidence that the system may be better described as a pair of
non-interacting inequivalent dimers, \textit{i.e} double dimers. However, the
existence of long-range magnetic order below $T_c \approx 28$~K immediately
questions such a description.  Although no evidence of the lozenge model is
observed, future studies on single crystals may further clarify the appropriate
magnetic Hamiltonian.
\end{abstract}

\maketitle

\section{Introduction}

\begin{figure}
\includegraphics[width=3.0in]{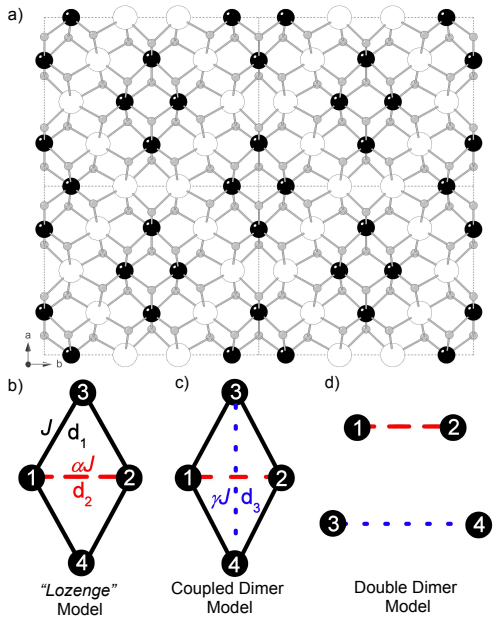}
\caption{Crystal structure and potential exchange interactions for Na$_3$RuO$_4$.
(a) Crystal structure as viewed along the $c$-axis showing a single plane of atomic
sites for $\pm \frac{c}{4}$ unit cells.  Ru, O and Na sites are black, gray, and
white respectively.  The monoclinic $C$~$2/m$ unit cell is shown as dashed
grey lines.   (b)-(d) are three possible models of exchange interactions between
Ru sites in the $ab$ plane as discussed in the text. The double dimer model is
not spatially confined, and represents two individual dimers.}
\label{NaRuOStructure}
\end{figure}

Magnetic materials have received continuous research interest since the initial
description of the Heisenberg Hamiltonian \cite{Heis26,Gatt06,Dag96,Kah93}.
This is due to interest in both the possible technological impact\cite{Nie00}, as
well as, fundamental physical phenomenon that many such
materials display \cite{Bar99}. Perovskite-based alkali metal ruthenates have just
recently started gaining attention\cite{sha80,shi04a,shi04b,regan2005}.  The
ruthenates exhibit a range of properties from ferro- and para-magnetism to
 superconductivity \cite{cal66,Rij99,Mae94}, and have been
shown to demonstrate an interesting cross-road in condensed matter physics \cite{cav04,Reg06,Alex03,Mog04}. The Na-Ru-O system has sparked interest into the many different analogs which present various magnetic properties from short magnetic order to paramagnetic behavior.\cite{regan2005} Na$_{3}$RuO$_{4}$ is one such analog that has induced similar queries over it's magnetic structure.

The structure of Na$_{3}$RuO$_{4}$ was first examined by Darriet \textit{et al.}
 \cite{Darr74}. The refinement of the crystal structure in these preliminary
measurements show that Na$_{3}$RuO$_{4}$ consists of oxygen coordinated
sodium and ruthenium sites within the $ab$ plane, separated by a single layer
of sodium sites displaced along the $c$-axis.  The structure of Na$_{3}$RuO$_{4}$
was recently re-refined and was determined to be mono-clinic, with space group
$C2$/$m$ and lattice parameters  $a = 11.0295(6)$~\AA, $b = 12.8205(7)$~\AA,
$c = 5.7028(3)$~\AA, and $\beta=109.90(3)^{\circ}$\cite{regan2005,regannote}.
Figure~\ref{NaRuOStructure}(a) illustrates a single plane of Ru ions together with
coplanar oxygen and sodium ions.  The Ru ions are octahedrally coordinated
through shared oxygens, and each Ru$^{+5}$ ion may be modeled as having
a local spin $S=\frac{3}{2}$.  This arrangement of ions suggest a local tetramer
or lozenge spin system as shown in Fig.~\ref{NaRuOStructure}(a)-(b).  An isolated spin-lozenge with exchange constants $J=3.36$~meV and
$\alpha J=3.88$~meV was first proposed by Drillon $et~al.$ in order
to describe magnetic susceptibility measurements on this material\cite{Dril77}.
However, the existence of  long range antiferromagnetic order below $T\approx 30$~K
was established using Mossbauer spectroscopy \cite{gib80}, and provided the first indication that the suggestion of antiferromagnetic
tetramer clusters in
Na$_{3}$RuO$_{4}$ may be incorrect.  Recent measurements of magnetic susceptibility
have also been interpreted in terms of a spin-tetramer model\cite{regan2005}.
 Temperature dependent neutron diffraction studies have confirmed the existence
 of long-range magnetic order below $T\approx 30$~K.  This
 long-range order immediately calls into question the accuracy of an isolated
 spin tetramer model of Na$_{3}$RuO$_{4}$. 
 
In the following sections, we present the exact analytical solutions for the energy eigenstates of a general Heisenberg spin S tetramer, and then apply these calculations to the case of spin-3/2 to determine the exact zero-field magnetic susceptibility and inelastic neutron scattering intensities. Through an examination of 
 thermodynamic and spectroscopic properties of Na$_3$RuO$_4$ and a
 comparison with theoretical predictions for various isolated spin tetramers, we shall see
 that the isolated tetramer model is indeed inappropriate  for Na$_{3}$RuO$_{4}$.

\section{Spin-S Coupled Dimer Model}

Figure~\ref{NaRuOStructure}(b)-(d) shows the individual configurations of the
coupled dimer models we examined.  As discussed above, Na$_{3}$RuO$_{4}$
is suggested to consist of isolated Ru$^{5+}$ (spin-3/2) tetramer clusters, where
the tetramers are in a lozenge configuration.   This is illustrated in
Fig.~\ref{NaRuOStructure}(b).  In this case, four Ru$^{5+}$ ions have
super-exchange interactions through Ru-O-Ru bonds, where the bond lengths were
determined by neutron diffraction and are given as 3.20, 3.20 and 5.56~\AA~ for distances
$d_1$, $d_2$, and $d_3$, respectively, as illustrated in
Fig.~\ref{NaRuOStructure}~\cite{regan2005}.  A more general case of the
spin-lozenge model includes a non-zero exchange interaction $\gamma J$
resulting in the coupled dimer configuration, Fig.~\ref{NaRuOStructure}(c).
If $J = 0$,  one recovers two isolated dimers or a double dimer configuration,
\textit{c.f.} Fig.~\ref{NaRuOStructure}(d).  We note that the double dimer model
is not spatially confined to the four Ru$^{5+}$ ions in the lozenge configuration,
but could also represent other dimer interactions in the Na$_{3}$RuO$_{4}$
crystal structure.  Using the coupled dimer model Hamiltonian, we determine
the eigenstates for general $S$, and calculate the corresponding magnetic
susceptibility for fitting purposes.  Then, with the choice of the appropriate magnetic
ground state, the excitations observed with inelastic neutron scattering (INS)
and their corresponding structure factors are also determined.

\subsection{Hamiltonian and Energy Eigenstates}

All three of the dimer configurations in Fig.~\ref{NaRuOStructure} can be described
by a single Hamiltonian, Eq. \ref{H_lozenge}.  By using this model, we can clearly examine the three
possible configurations that may describe Na$_{3}$RuO$_{4}$.  Using
nearest-neighbor Heisenberg interactions and a Zeeman magnetic field term for
magnetic fields $\mathbf{B}$ defining the $z$-axis, the general Hamiltonian is
\be
\begin{array}{c}
{\cal H} =
{J }
\Big[ \big(\,
\vec{\rm S}_{1}\cdot\vec{\rm S}_{3} +
\vec{\rm S}_{1}\cdot\vec{\rm S}_{4} +
\vec{\rm S}_{2}\cdot\vec{\rm S}_{3} +
\vec{\rm S}_{2}\cdot\vec{\rm S}_{4}\,
\big)\\
+\; \alpha\,
\vec {\rm S}_{1}\cdot\vec {\rm S}_{2} +\gamma\,
\vec {\rm S}_{3}\cdot\vec {\rm S}_{4} \Big]
- \big(S^z_{1} +S^z_{2}+S^z_{3}+S^z_{4} \big) g \mu_B B,\\
\end{array}
\label{H_lozenge}
\ee
where $\alpha J$ is the interaction for the $\alpha$-dimer, $\gamma J$ is the interaction for the $\gamma$-dimer, and $\mu_B$ is
the Bohr magneton.  We define the exchange interaction as positive for
antiferromagnetic interactions, and $\vec {\rm S}_i$ is the quantum spin operator
for a spin-$S$ ion at site $i$=1,2,3,4.  The Zeeman term interacts with the
$z$-component of the spin Hamiltonian, lifting the degeneracy of magnetic
substates in applied magnetic fields.

The Hamiltonian, Eq.~\ref{H_lozenge}, is rotationally invariant in spin space,
such that the total spin, $S_{tot}$, and $S_{z}$ are good quantum numbers.
For the general case of a spin-$S$ tetramer cluster, the energy eigenstates
have the total spin decompositions given by
\be
\begin{array}{c}
\prod_{S_{tot}=1}^{4} S_n = \sum_{S_{tot}=0}^{2S}(4S-S_{tot})^{\left(\frac{(S_{tot}+1)(S_{tot}+2)}{2} \right)} \oplus \\ \\
\sum_{S_{tot}=0}^{2S-1}S_{tot}^{\frac{1}{2}(4S+2+8S_{tot}S+S_{tot}-3S_{tot}^2)},\\
\end{array}
\ee
where the total number of magnetic states in a general $S$ tetramer are
$(2S+1)^4$. Therefore for the $S=3/2$ tetramer, the energy eigenstates
have S$^4$ = 256 magnetic states and the spin decomposition is given
by
\be
\begin{array}{c}
\frac{3}{2} \otimes \frac{3}{2} \otimes \frac{3}{2} \otimes \frac{3}{2}
= \\
\mathrm{Individual ~ Spins,}\\ \\
\big(3 \oplus 2 \oplus 1 \oplus 0 \big) \otimes \big(3 \oplus 2 \oplus 1 \oplus 0 \big) =\\
\mathrm{Dimer ~ States, and}\\ \\
6 \oplus 5^3 \oplus 4^6 \oplus 3^{10} \oplus 2^{11} \oplus 1^9 \oplus 0^4\\
\mathrm{Tetramer ~ States}.\\
\end{array}
\label{CGseries}
\ee
The superscript in the tetramer states denote multiple $S_{tot}$ states. Each
multiplet containing $2S_{tot}+1$ magnetic states, which are degenerate given
an isotropic
magnetic Hamiltonian such as the Heisenberg form of Eq. \ref{H_lozenge}, where
the degenerate states can be split by a magnetic field. This breakdown of the dimer
and tetramer spin states helps clarify which dimer states are interacting to create the composite tetramer states.

By expanding the Kambe approach \cite{Kam50,Gatt06}, we can rewrite the
Hamiltonian in terms of total spin for the individual diagonalizable components,
in which the eigenstates and eigenvalues of the Hamiltonian may be found by
 diagonalization in the convenient basis of two dimers. This approach gives
information about the states of the dimers as the tetramer states are determined,
which allows a clearer picture of the magnetic excitations. In practice, we employ
the usual set of $\hat z$-polarized magnetic basis states. The energy levels are
then determined simply by considering a dimer basis, where $S_{\alpha}$
corresponds to the spin state of the $\alpha$ dimer and $S_{\gamma}$ corresponds
to the spin state of a $\gamma$ dimer as described in the Hamiltonian,
Eq.~\ref{H_lozenge}.  Using this dimer basis, the energy levels for the
general $S$ coupled dimer can be determined exactly and are given
by
\be
E = \frac{J}{2} \big[ \mathscr{S}_{tot} + \mathscr{S}_{\alpha}  (\alpha - 1) +  \mathscr{S}_{\gamma} (\gamma - 1) - 2 (\alpha + \gamma) \mathscr{S} \big]
\ee
where  $\mathscr{S}_{tot} = S_{tot}(S_{tot} + 1)$ with $S_{tot}$ denoting the
magnetic state of the system, $\mathscr{S}_{\alpha}=S_{\alpha}(S_{\alpha}+1)$
and $S_{\alpha}$ is the spin state of the $\alpha$ dimer ($S_{1}$-$S_{2}$ dimer),
$\mathscr{S}_{\gamma}=S_{\gamma}(S_{\gamma}+1) $ and $S_{\gamma}$ is the
spin state of the $\gamma$ dimer ($S_{3}$-$S_{4}$ dimer), and
$\mathscr{S} = S(S+1)$ with $S$ being the spin of the system.

\begin{table}
\caption{\textbf{Energy levels for a spin-$\frac{3}{2}$ lozenge}}
\begin{ruledtabular}
\begin{tabular}{lcc}
 \textbf{S$_{tot}$} & \textbf{Spin State$^a$}  & \textbf{Energy}  \\
 & \textbf{$|$ S$_{tot}$ S$^z_{tot}$ $>_{S_{\alpha},S_{\gamma}}$}& \textbf{Level$^b$} \\
\hline \hline \\
6 & $|$ 6 S$^z_{tot}$ $>_{3,3}$& $J(9+\frac{9}{4}\alpha+\frac{9}{4} \gamma)$ \\
\\
\hline \\
5 & $|$ 5 S$^z_{tot}$ $>_{3,3}$& $J(3+\frac{9}{4}\alpha+\frac{9}{4} \gamma)$ \\

 & $|$ 5 S$^z_{tot}$ $>_{3,2}$& $J(6+\frac{9}{4}\alpha-\frac{3}{4} \gamma)$ \\

 & $|$ 5 S$^z_{tot}$ $>_{2,3}$& $J(6-\frac{3}{4}\alpha+\frac{9}{4} \gamma)$ \\
\\
\hline \\
4 & $|$ 4 S$^z_{tot}$ $>_{3,3}$& $J(-2+\frac{9 }{4}\alpha+\frac{9}{4} \gamma)$ \\

 & $|$ 4 S$^z_{tot}$ $>_{3,2}$& $J(1+\frac{9}{4}\alpha-\frac{3}{4} \gamma)$ \\

 & $|$ 4 S$^z_{tot}$ $>_{3,1}$& $J(3+\frac{9 }{4}\alpha-\frac{11}{4} \gamma)$ \\

 & $|$ 4 S$^z_{tot}$ $>_{2,3}$& $J(1-\frac{3 }{4}\alpha+\frac{9}{4} \gamma)$ \\

 & $|$ 4 S$^z_{tot}$ $>_{1,3}$& $J(3-\frac{11}{4}\alpha+\frac{9}{4} \gamma)$ \\

 & $|$ 4 S$^z_{tot}$ $>_{2,2}$& $J(4-\frac{3}{4}\alpha-\frac{3}{4} \gamma)$ \\
\\
\hline \\
3 & $|$ 3 S$^z_{tot}$ $>_{3,3}$& $J(-6+\frac{9}{4}\alpha+\frac{9}{4} \gamma)$ \\

 & $|$ 3 S$^z_{tot}$ $>_{3,2}$& $J(-3+\frac{9}{4}\alpha-\frac{3}{4} \gamma)$ \\

 & $|$ 3 S$^z_{tot}$ $>_{3,1}$& $J(-1+\frac{9}{4}\alpha-\frac{11}{4} \gamma)$ \\

 & $|$ 3 S$^z_{tot}$ $>_{2,3}$& $J(-3-\frac{3}{4}\alpha+\frac{9}{4} \gamma)$ \\

 & $|$ 3 S$^z_{tot}$ $>_{1,3}$& $J(-1-\frac{11}{4}\alpha+\frac{9}{4} \gamma)$ \\

 & $|$ 3 S$^z_{tot}$ $>_{2,2}$& $J(-\frac{3}{4} \alpha-\frac{3}{4} \gamma)$ \\

 & $|$ 3 S$^z_{tot}$ $>_{2,1}$& $J(2-\frac{3}{4}\alpha-\frac{11}{4} \gamma)$ \\

 & $|$ 3 S$^z_{tot}$ $>_{1,2}$& $J(2-\frac{11}{4}\alpha-\frac{3}{4} \gamma)$ \\

 & $|$ 3 S$^z_{tot}$ $>_{3,0}$& $J(\frac{9}{4} \alpha-\frac{15}{4} \gamma)$ \\

 & $|$ 3 S$^z_{tot}$ $>_{0,3}$& $J(-\frac{15 }{4}\alpha+\frac{9}{4} \gamma)$ \\
\\
\hline \\
2 & $|$ 2 S$^z_{tot}$ $>_{3,3}$& $J(-9+\frac{9}{4}\alpha+\frac{9}{4} \gamma)$ \\

 & $|$ 2 S$^z_{tot}$ $>_{3,2}$& $J(-6+\frac{9}{4}\alpha-\frac{3}{4} \gamma)$ \\

 & $|$ 2 S$^z_{tot}$ $>_{3,1}$& $J(-4+\frac{9}{4}\alpha-\frac{11}{4} \gamma)$ \\

 & $|$ 2 S$^z_{tot}$ $>_{2,3}$& $J(-6-\frac{3}{4}\alpha+\frac{9}{4} \gamma)$ \\

 & $|$ 2 S$^z_{tot}$ $>_{1,3}$& $J(-4-\frac{11}{4}\alpha+\frac{9}{4} \gamma)$ \\

 & $|$ 2 S$^z_{tot}$ $>_{2,2}$& $J(-3-\frac{3}{4}\alpha-\frac{3}{4} \gamma)$ \\

 & $|$ 2 S$^z_{tot}$ $>_{2,1}$& $J(-1-\frac{3}{4}\alpha-\frac{11}{4} \gamma)$ \\

 & $|$ 2 S$^z_{tot}$ $>_{1,2}$& $J(-1-\frac{11}{4}\alpha-\frac{3}{4} \gamma)$ \\

 & $|$ 2 S$^z_{tot}$ $>_{1,1}$& $J(1-\frac{11}{4}\alpha-\frac{11}{4} \gamma)$ \\

 & $|$ 2 S$^z_{tot}$ $>_{2,0}$& $J(-\frac{3 }{4}\alpha-\frac{15}{4} \gamma)$ \\

 & $|$ 2 S$^z_{tot}$ $>_{0,2}$& $J(-\frac{15}{4} \alpha-\frac{3}{4} \gamma)$ \\
\\
\hline \\
1 & $|$ 1 S$^z_{tot}$ $>_{3,3}$& $J(-11+\frac{9}{4} \alpha+\frac{9}{4} \gamma)$ \\

 & $|$ 1 S$^z_{tot}$ $>_{3,2}$& $J(-8+\frac{9}{4}\alpha-\frac{3}{4} \gamma)$ \\

 & $|$ 1 S$^z_{tot}$ $>_{2,3}$& $J(-8-\frac{3 }{4}\alpha+\frac{9}{4} \gamma)$ \\

 & $|$ 1 S$^z_{tot}$ $>_{2,2}$& $J(-5-\frac{3}{4} \alpha-\frac{3}{4} \gamma)$ \\

 & $|$ 1 S$^z_{tot}$ $>_{2,1}$& $J(-3-\frac{3}{4}\alpha-\frac{11}{4} \gamma)$ \\

 & $|$ 1 S$^z_{tot}$ $>_{1,2}$& $J(-3-\frac{11}{4}\alpha-\frac{3}{4} \gamma)$ \\

 & $|$ 1 S$^z_{tot}$ $>_{1,1}$& $J(-1-\frac{11}{4}\alpha-\frac{11}{4} \gamma)$ \\

 & $|$ 1 S$^z_{tot}$ $>_{1,0}$& $J(-\frac{11}{4}\alpha-\frac{15}{4} \gamma)$ \\

 & $|$ 1 S$^z_{tot}$ $>_{0,1}$& $J(-\frac{15 }{4}\alpha-\frac{11}{4} \gamma)$ \\
\\
\hline\\
0 & $|$ 0 S$^z_{tot}$ $>_{3,3}$& $J(-12+\frac{9}{4} \alpha+\frac{9 }{4}\gamma)$ \\

 & $|$ 0 S$^z_{tot}$ $>_{2,2}$& $J(-6-\frac{3}{4} \alpha-\frac{3}{4} \gamma)$ \\

 & $|$ 0 S$^z_{tot}$ $>_{1,1}$& $J(-2-\frac{11 }{4}\alpha-\frac{11 }{4}\gamma)$ \\

 & $|$ 0 S$^z_{tot}$ $>_{0,0}$& $J(-\frac{15 }{4}\alpha-\frac{15 }{4}\gamma)$ \\
\\
\end{tabular}
\end{ruledtabular}
\label{coupling_table}

\footnotemark{$S_{\alpha}$ and $S_{\gamma}$ are the spin state of the $\alpha$ and $\gamma$ dimers.}

\footnotemark{The magnetic field splitting of the states is given by the addition of a $S^z_{tot} g \mu_B B$ to the energy of the magnetic states.}

\end{table}

For the $S = \frac{3}{2}$ tetramer, the energy eigenstates of the isotropic
Heisenberg Hamiltonian with general $S^z_{tot}$ are given in
Table \ref{coupling_table}. The magnetic substates are all degenerate in the
absence of an applied magnetic field, but can be split linearly in accordance
with the Zeeman field term. The individual states can be split into $2S_{tot}+1$
states with magnetic field. Therefore, the energy levels split by $g \mu_B B S^z_{tot}$,
where $S^z_{tot}$ is from $0\ldots(2S_{tot}+1)$. The ground state of the $S = \frac{3}{2}$
tetramer can either have a nonmagnetic $S=0$ ground state or a magnetic $S=1$
ground state depending upon the values of $\alpha$ and $\gamma$.   Simple
examination of the energy levels in Table~\ref{coupling_table} indicates that,
assuming antiferromagnetic exchange for $J$, the ground state will be
non-magnetic when both $\alpha$ and $\gamma$ are less than $\frac{4}{3}$.

\subsection{Spin-$\frac{3}{2}$ Magnetic Observables}
Magnetic observables, \textit{i.e.} specific heat and magnetic susceptibility, associated
with the general coupled dimer model can be derived via the partition function
and eigenvalues.  Thus, such macroscopic measurements of Na$_{3}$RuO$_{4}$
may serve to place limits on the nature of the interactions.  We also determine
the excitation energies and structure factors which would be observed in
INS measurements.  Both thermodynamic and spectroscopic measurements
should be consistent for any appropriate description of the experimental system.

\subsubsection{Magnetic Susceptibility and Heat Capacity}

 We now present the method for determining the partition function and
 magnetic susceptibility for the $S=3/2$ coupled dimer model. Due to the length of
the equations, we present the eigenstates and eigenvalues explicitly in
Table~\ref{coupling_table} and represent the magnetic observables as
summations over energy eigenvalues.  Using this method, the canonical
partition function is given by
 \be
Z = \sum_{i=1}^N\, e^{- \beta \E_i}
 = \sum_{\E_i}\, (2{\rm S}_{tot}+1)\, e^{- \beta \E_{i}} \
\label{partition}
\ee
\noindent and the magnetic susceptibility is given by
\be
\begin{array}{c}
\chi = \frac{\beta}{Z}
\sum_{i=1}^N  \, (M_z^2)_i\, e^{-\beta \E_i}\\ \\
=  \frac{1}{3} (g \mu_B)^2 \frac{\beta}{Z}
\sum_{\E_i} \, (2{\rm S}_{tot}+1) \, ({\rm S}_{tot}+1)\,
{\rm S}_{tot}\, e^{- \beta \E_i} \ .\\
\end{array}
\label{mag_suscept}
\ee
In these formulas, the sum $\sum_{i=1}^N$ is over all $N$ independent energy
eigenstates (including magnetic substates), the sum $\sum_{\E_i}$ is over
energy
levels only, $M_z = m g \mu_B$ where
$m = \S_{tot}^z/\hbar $ is the integral or half-integral
magnetic quantum number, $g$ is the Lande $g$-factor, $\beta=\frac{1}{k_B T}$ and
$k_B$ is Boltzmann's constant \cite{Kah93}. The heat capacity can also be determined from
\be
C =  k_B \beta^{\, 2}\, \frac{d^2 \! \ln (Z)}{d \beta^2} \ .
\label{C}
\ee
In general, the heat capacity is especially useful for confirming the proper accounting
of eigenstates through a numerical calculation of the magnetic entropy of the tetramer system at
large temperature,
\bd
\frac{S}{k_B} =
\int_0^{\infty}\!\! \frac{C}{k_B} \; \frac{d\beta}{\beta} =
\ed
\be
\ln({\cal N}/{\cal N}_0) \ =
\begin{cases}
2\ln(2), &           \text{$\alpha  < \frac{4}{3}~ \mathrm{and} ~\gamma < \frac{4}{3}$} \\
\ln(\frac{256}{3}), &           \text{$\alpha > \frac{4}{3}~ \mathrm{or}  ~\gamma > \frac{4}{3} $}.\\
\end{cases}
\ee
Here ${\cal N}$ is the dimensionality of the full Hilbert space
and ${\cal N}_0$ is the degeneracy of the ground state manifold. The numerical calculation of entropy within various regions of $\alpha$ and $\gamma$ confirms the ground state of the spin-3/2 tetramer.

\subsubsection{Inelastic Neutron Scattering}

The experimental focus of this paper is the use of INS to investigate the nature of
the magnetic interactions and their respective excitations. Using methods presented
in Haraldsen $et~al.$\cite{Squ78,Har05}, we next determine the excitation energies
and structure factors for the observable transitions of the coupled dimer models shown
in Figs.~\ref{NaRuOStructure}(b)-(d).

For transitions out of the ground state, the excitation energy, $\hbar \omega$, is simply
the difference in energy between the excited and ground states.  Such excitations
would be non-dispersive in the absence of inter-tetramer exchange.  The differential
cross-section of finite systems is proportional to the neutron scattering structure
factor
\be
S(\vec q\, ) =
|{\rm F}(\vec q\,)|^2 \left[ \sum_{\lambda_f}\
 \ \langle \Psi_f (\lambda_f)|
V_a
| \Psi_i \rangle^2  \  \right]
\label{Sab_def}
\ee
where ${\rm F}(\vec q\,)$ is the magnetic form factor and the vector $V_a(\vec q\,) $
is a sum of spin operators over all magnetic ions in a unit cell
\be
V_a = \sum_{{\vec x}_i} {\S}_a(\vec x_i)\;
e^{i\vec q \cdot \vec x_i } \ .
\label{Va_defn}
\ee
For rotationally invariant magnetic interactions
and an $S_{tot} = 0$ ground state in the $T=0$ limit, only
$S_{tot}~=~1$ final states as shown in Table \ref{coupling_table} are observable
via INS. However, due to the nature of the tetramer states as being composite dimer
states, this implies that it is only possible to excite transitions of the individual dimers
which make up the tetramer structure, $\Delta S_{\alpha / \gamma} = \pm 1, 0$.
To interpret neutron experiments on powder samples,
we require an orientation average of the unpolarized
single-crystal neutron scattering structure factor.
We define this powder average by
\be
{\bar S}(q) = \int \frac{d\Omega_{\hat q}}{4\pi}\, S(\vec q\, ) \ .
\label{Strpowavg}
\ee

With respect to the spin-3/2 rhombus model, the values of the magnetic interactions
quoted in the literature suggest a $S_{tot} = 0$ ground state, with dimer spins
$S_{\alpha} = 3$ and $S_{\gamma} = 3$\cite{Dril77,regan2005}. Therefore, due to this selective restriction of the spin excitations, only three of
the nine $S_{tot} = 1$ states are accessible from that ground state through INS.
The respective excitation energies ($E_{S_{tot},S_{\alpha},S_{\gamma}}$) are
\be
\begin{array}{c}
E_{0,3,3 \rightarrow 1,3,3}  = J, \\
\\
E_{0,3,3 \rightarrow 1,3,2} = J(4-3\gamma), \\
\\
E_{0,3,3 \rightarrow 1,2,3} = J(4-3\alpha), \\
\end{array}
\ee
and the powder average INS structure factors ($\bar S(q)_{S_{\mathrm{tot}},S_{\alpha},S_{\gamma}}$) for these transitions are
\be
\begin{array}{c}
\bar S(q)_{0,3,3 \rightarrow 1,3,3}  = \\2 |{\rm F}(\vec q\,)|^2 (2-4j_0(qd_1)+j_0(qd_2)+j_0(qd_3)), \\
\\
\bar S(q)_{0,3,3 \rightarrow 1,3,2} = \frac{|{\rm F}(\vec q\,)|^2}{2}(1-j_0(qd_3)), \\
\\
\bar S(q)_{0,3,3 \rightarrow 1,2,3} = \frac{|{\rm F}(\vec q\,)|^2}{2}(1-j_0(qd_2)), \\
\label{eq:powderintensities}
\end{array}
\ee
where $d_1$, $d_2$, and $d_3$ are the interatomic separations (shown in
Fig.~\ref{NaRuOStructure}), $j_0(x)=\frac{sin(x)}{x}$, and $ |{\rm F}(\vec q\,)|$
is the Ru$^{5+}$ magnetic form factor (a parameterization is given by Parkinson $et~al.$\cite{Park03,Park05}).
The transition of $|00 >_{3,3} \rightarrow |1 S_{tot}^{z} >_{3,3}$ is an excitation of the full tetramer, while the other two transitions are excitations of individual dimers. This shows that out of the
nine possible spin-1 states to be excited for the spin-$\frac{3}{2}$ model, only three are accessible
by INS and will have a corresponding intensity profile. The unseen transitions are inaccessible because they require multiple spin transitions, and since a neutron can only provide one transition, only transitions that excite the individual components will be observed by neutron scattering.

\section{Experimental Techniques}

Powder samples of Na$_3$RuO$_4$ were prepared by solid-state reactions
from stoichiometric amount of NaOH and RuO$_2$.  The starting stoichiometric
mixture was initially ground together and then held at 500 $^{\circ}$C for 20 hr.
under an O$_2$ atmosphere.  After re-grinding, the powder was heated to 650
$^{\circ}$C for another 20 hr, again under an O$_2$ atmosphere.  The resulting
dark grey powder was reground and checked for impurity phases using
X-ray
powder diffraction.  If any impurity phases were evident,
the powder was refired and the process repeated.
This growth
procedure is similar to that described
in Ref.~\onlinecite{regan2005}.
Powder refinement of room temperature
X-ray diffraction measurements yielded lattice
parameters of $a=11.012(7)$, $b=12.809(9)$, $c=5.687(3)$~\AA, and $\beta=109.91(3)^{\circ}$ for the $C2/m$ monoclinic
unit cell\cite{regannote}.  These
values compare well to the fully refined structure
described in Ref.~\onlinecite{regan2005}.
Single crystals of appropriate mass are unfortunately not
yet available for INS measurements.

Heat capacity measurements were performed on
a small single crystal of mass $\approx 10$~mg, which was obtained through the
synthesis procedure described above.  This single
crystal grew as a small platelet, with
the $c$-axis normal to the plane of the platelet.
Heat capacity measurements were performed with a commercial calorimeter
between
$T=1.8$ K and 300~K, using the relaxation technique. Measurements
were carried out in zero and 8~T applied
magnetic fields, with the field applied along the $c$-axis of the
single crystal sample.

Magnetization measurements were
performed on powder and single crystal samples
using a commercial SQUID, as a function of applied magnetic field and
temperature.  SQUID measurements on the same single crystal sample
that was
used for heat capacity measurements agree well with those taken on a
powder sample.

INS measurements were performed
using the MARI time-of-flight spectrometer at the ISIS neutron
scattering facility \cite{mari}.  The sample consisted of $\approx 45$~g of
Na$_3$RuO$_4$ powder in a square aluminum foil sachet (approximately
50 by 50 by 8 mm), suspended
from the cold-tip of a closed-cycle He$^4$ refrigerator.
The sachet was oriented with
the 50x50 mm surface normal to the incident neutron beam.
An incident energy of $E_i = 25$ meV was used, and data
were taken at several temperatures between
$T=8$~K and $T=305$~K.   This configuration
resulted in a measured
instrumental energy resolution at the elastic position of
$\delta \hbar \omega = 0.982(7)$~meV full width at
half maximum (FWHM).  Data
were corrected for detector
sensitivity through room temperature
measurements on a vanadium standard.

\begin{figure}
\includegraphics[width=3.25in]{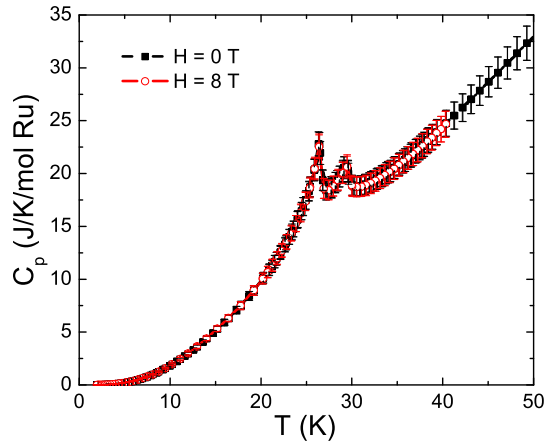}
\caption{Heat capacity of a Na$_3$RuO$_4$ single crystal.  Measurements
were performed at zero field (black squares) and $H = 8$~T (red open circles),
with $H\parallel c$. Error bars represent an estimated five percent error in the measurement.}
\label{Cp}
\end{figure}

INS measurements were also carried out using the HB3 triple-axis spectrometer
at the high flux isotope reactor (HFIR)
at Oak Ridge National Laboratory.
For these measurements, the sample consisted of 20.7 g of Na$_3$RuO$_4$
powder
in a cylindrical aluminum sample can of 18 mm diameter and 57 mm height.
The sample can was sealed
under He gas and mounted to the cold-tip of a
closed-cycle He$^4$ refrigerator. Horizontal
collimation was chosen as
$48^{\prime}-40^{\prime}-40^{\prime}-120^{\prime}$
between source
and monochromator, monochromator and sample, sample
and analyzer, and analyzer and detector, respectively. The
spectrometer was operated with fixed final energy,
$E_f =14.7$~meV, using a pyrolytic graphite (PG 002)
monochromator and analyzer. Pyrolytic graphite filters were placed
after the sample to substantially reduce higher-order spurious
scattering processes. In this configuration, the energy resolution
at the elastic position was $\delta\hbar\omega =1.10(2)$~meV FWHM, as
measured from the
incoherent scattering at $Q=1.2$~\AA$^{-1}$. The
wave vector resolution was measured to be $\delta Q = 0.0407(7)$~
\AA$^{-1}$ FWHM using the (110) nuclear Bragg peak.  All
measurements were made for fixed incident neutron monitor count.

\begin{figure}
\includegraphics[width=4.0in]{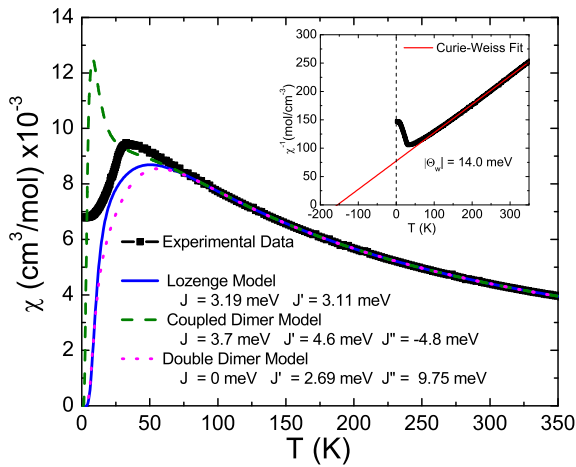}
\caption{Magnetic susceptibility of Na$_3$RuO$_4$ powder (black squares),
showing fits using the three models: Lozenge (solid blue), coupled dimer
(dashed green), double dimer (dotted magneta). The inset shows $\chi ^{-1}$,
with a Curie-Weiss fit as described in the text (which gives $|\Theta_{CW}|$
= 14.0 meV, and is consistent with dominantly antiferromagnetic interactions.)}
\label{ChivsT}
\end{figure}

Elastic neutron scattering measurements were also
performed using the HB3 triple-axis spectrometer, with
$E_i = E_f = 14.7$~meV.  These measurements
were performed on the same powder sample as the
inelastic HB3 measurements, with horizontal
collimation $48^{\prime}-20^{\prime}-20^{\prime}-70^{\prime}$.
This resulted in an energy resolution
at the elastic position of $\delta\hbar\omega \approx 0.8$~meV
FWHM. The wave vector resolution was measured
to be $\delta Q = 0.0254(9)$~
\AA$^{-1}$ FWHM using the (110) nuclear Bragg peak.

INS measurements were also performed to place limits on the
value of a possible energy gap in the excitation spectrum.  These
were performed using the IRIS backscattering spectrometer at the ISIS neutron source at the Rutherford Appleton Laboratory \cite{iris}. The sample
measured was the identical powder used for the HB3 measurements.  The IRIS spectrometer was operated at 25~Hz with cooled PG002 analyzers ($T=10$~K) and a Berylium filter ($T=25$~K) to avoid contamination from higher-order reflections, resulting in a $17.5$~$\mu$eV FWHM energy resolution at the elastic position as measured with a vanadium standard. The high-resolution (back-scattering) diffraction banks with a resolution of   $\frac{\Delta Q}{Q} = 2.5$ E-3 were also used on IRIS.

\section{Experimental Results}

\begin{figure}
\includegraphics[width=3.0in]{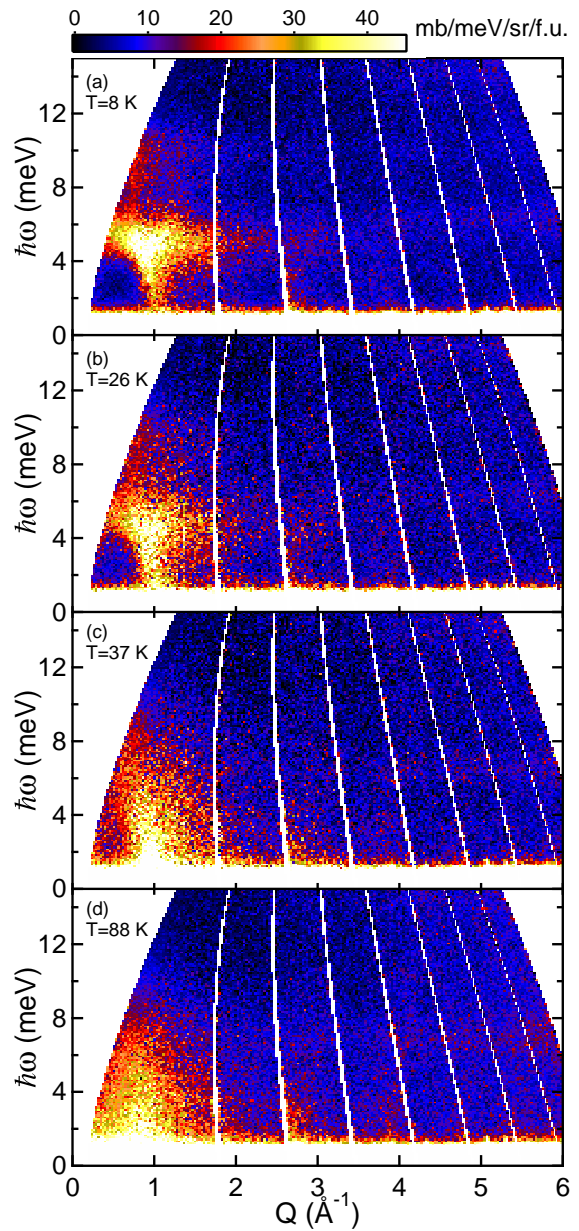}
\caption{INS intensity from Na$_3$RuO$_4$ powder versus energy and
momentum transfer at (a) $T = 8$~K, (b) $T=26$~K, (c) $T=37$~K and (d)
$T=88$~K.  These measurements were carried out on the MARI spectrometer at an incident energy of $E_i = 25$~meV.}
\label{MARITemp}
\end{figure}

Figure~\ref{Cp} shows the heat capacity as a function of temperature for $H = 0$~and
8 T.  There are two clear lambda-like anomalies at $T\approx23$ and
$T\approx28$~K, signifying phase transitions at these temperatures.
Previous neutron diffraction measurements have shown the existence of
only a single, broad phase transition near 30 K in Na$_3$RuO$_4$,
corresponding to the onset of long-range magnetic order\cite{regan2005}.
Our heat capacity measurements indicate that the observed broad transition is
likely due to two successive transitions that occur at similar temperatures. No
change was noted in these transition temperatures when measured at $H=0$
and $H=8$ T.

The magnetic susceptibility of Na$_3$RuO$_4$ was measured over the range
$2\leq T \leq 350$~K; the resulting data is shown in Fig.~\ref{ChivsT}. The
susceptibility also shows evidence for a phase transition near $T\approx 30$ K.
The negative intercept in $\chi^{-1}(T)$ and the decrease in $\chi(T)$ below the
transition temperature are consistent with dominantly antiferromagnetic interactions.

Figure \ref{MARITemp} shows the temperature dependent inelastic neutron scattering
data taken on the MARI spectrometer at ISIS.  There is significant inelastic
scattering intensity in the vicinity of $\hbar\omega\approx 5$~meV, which decreases
in intensity rapidly with increasing wave-vector.  The wave-vector dependence implies
that the scattering is magnetic in origin.  In the $T=8$~K data, a weak excitation near
$\hbar\omega\approx 10$~meV is also evident.  As the temperature increases,
the inelastic scattering intensity rapidly decreases and moves to smaller
wave-vectors, consistent with an evolution from antiferromagnetic spin-waves
to paramagnetic scattering with increasing temperature.  We speculate that the
excitations observed below $T_N$ are acoustic and optical spin-waves associated
with the long-range ordered phase.  Higher incident energy measurements were
also performed, which show evidence for phonon excitations above 20 meV.

\begin{figure}
\includegraphics[width=3.75in]{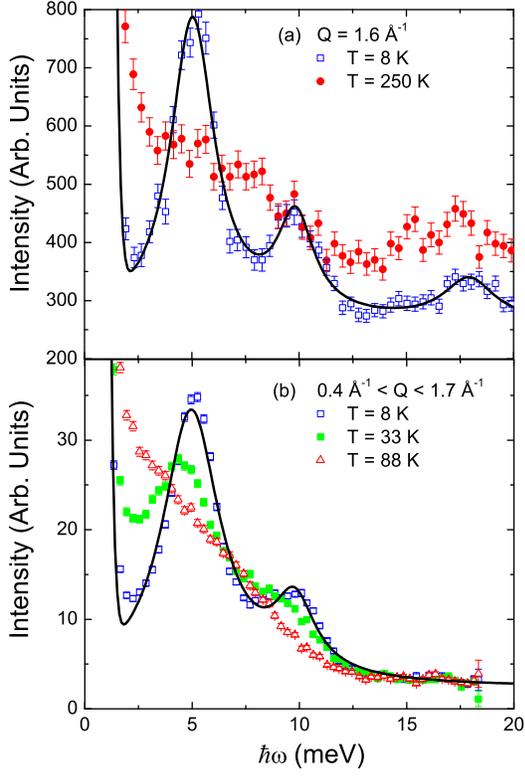}
\caption{INS intensity from Na$_3$RuO$_4$ powder versus energy transfer at
(a) $Q=1.6$~\AA$^{-1}$ at $T = 8$ and $250$~K and (b) integrated between
$0.4$~\AA$^{-1} < Q < 1.6$~\AA$^{-1}$ at $T = 8$~K, $30$~K and $88$~K.  Black
lines are Lorentzian fits, as described in the text.  The data in (a) are from HB3, and
the data in (b) are from MARI.}
\label{IvsE}
\end{figure}

In Fig.~\ref{IvsE}, we show the scattering intensity as a function of energy transfer for
the single wave-vector $Q =1.6$~\AA$^{-1}$, measured on the HB3 spectrometer, as
well as the integrated scattering intensity for $ 0.4$~\AA$^{-1} < Q < 1.7$~\AA$^{-1}$
on the MARI spectrometer. Single Lorentzian fits to the low-temperature data
suggest modes at
$\hbar\omega = 5.03\pm0.08$, $9.8\pm0.2$ and $17.9\pm0.3$~meV
(for the data shown in
Fig.~\ref{IvsE}(a)), and $\hbar\omega = 4.95\pm0.04$ and $9.8\pm0.1$ meV for the
data shown in Fig.~\ref{IvsE}(b).  The increase in intensity of the 18~meV excitation
with increasing temperatures suggests that it is likely a phonon excitation.

\begin{figure}
\includegraphics[width=3.75in]{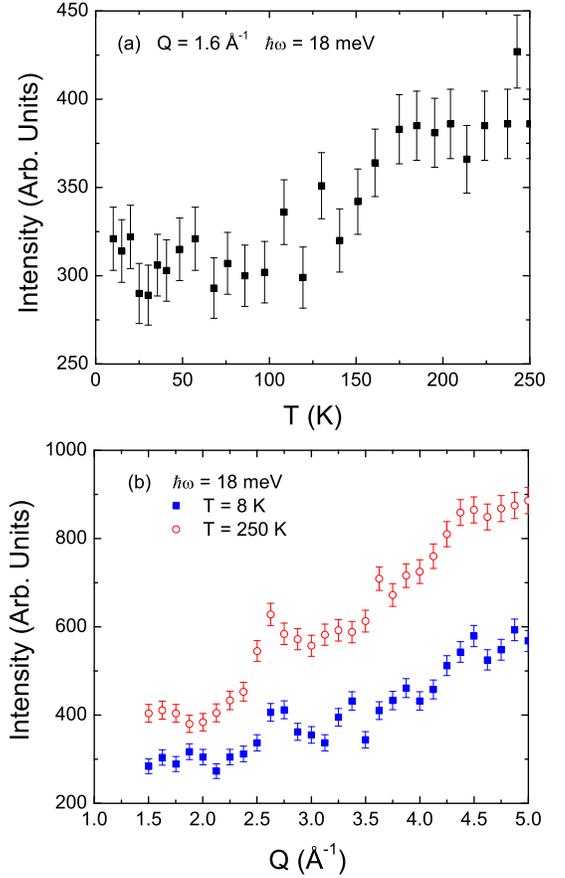}
\caption{(a) INS intensity versus temperature at $\hbar \omega = 18$~meV and
$Q = 1.6$~\AA$^{-1}$.  (b) INS intensity versus momentum transfer at
$\hbar \omega = 18$~meV  at $T = 8$ ~and $T = 250$~K.  Data were
acquired using HB3.}
\label{TempHigh}
\end{figure}

The 18~meV phonon excitation was investigated more carefully as a function of
temperature and wave-vector, as shown in Fig.~\ref{TempHigh}.  These data show
a monotonic increase in scattering intensity as the temperature is increased, and
an increase in scattering intensity with increasing wave-vector.  These results
further support the identification of the 18~meV mode with a phonon excitation.

\begin{figure}
\includegraphics[width=3.0in,height=2.5in]{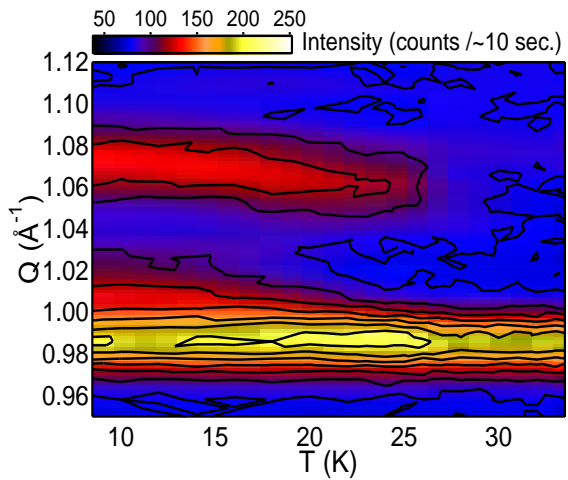}
\caption{Elastic scattering intensity of Na$_3$RuO$_4$ as function of momentum
transfer and temperature. Data were acquired using the HB3 spectrometer.  The
contour lines correspond to intervals of 20 counts per ten seconds, and are only
plotted for
count rates between 80 and 200 counts per ten seconds.
The data were obtained at 20 temperatures between $T=8.5$ and $34$~K.}
\label{braggcontour}
\end{figure}

\begin{figure}
\includegraphics[scale=0.8]{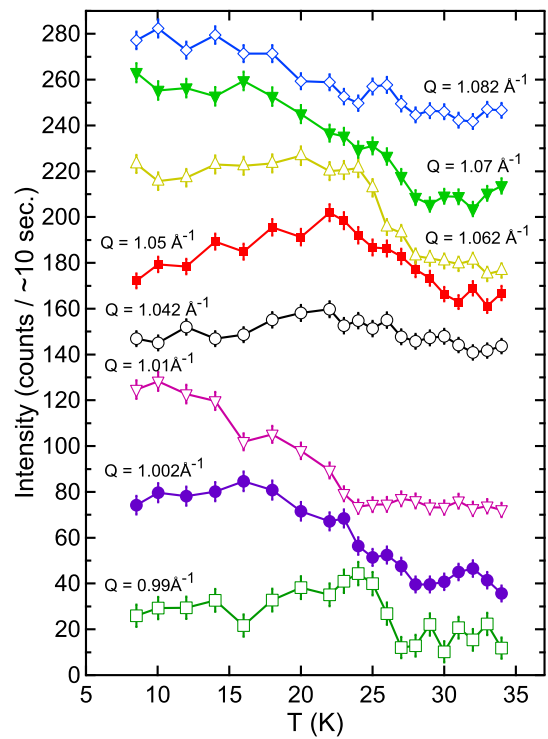}
\caption{Elastic scattering intensity from Na$_3$RuO$_4$ powder versus temperature
for several wave-vectors. The data correspond to the individual wave-vectors shown
in Fig.~\ref{braggcontour}.  The data have been offset along the vertical axis
for presentation.}
\label{braggtempdepend}
\end{figure}

We also examined the temperature dependence of the elastic scattering in the vicinity
of $Q \approx 1$~\AA$^{-1}$.  Figure~\ref{braggcontour} shows the scattering intensity
observed in the HB3 Na$_3$RuO$_4$ powder measurement as a function of
temperature and wave-vector. As the temperature is decreased, there is an increase
in the scattering intensity at $Q\approx 0.99$ and $\approx 1.07$~\AA$^{-1}$
corresponding to the transition to long range magnetic ordering.  Below
$T\approx 25$~K, the magnetic Bragg peaks appear to move as a function
of decreasing temperature.  This is also evident in Fig.~\ref{braggtempdepend},
which shows the scattering intensity as a function of temperature for various
individual momentum transfers. For certain Q values, the scattering intensity
shows non-monotonic temperature dependence, for example as shown in Fig. \ref{braggcontour} at $Q=0.99$~\AA$^{-1}$
and $1.05$~\AA$^{-1}$.  These behaviors may be due to the presence of two
magnetic phase transitions near 25-30~K as seen in the heat capacity measurements.

High resolution backscattering measurements investigated the magnetic spectrum for energy transfers below 1.7 meV (shown in Fig.~\ref{IRIS-HB3-figure}(a)). Excitations are populated out of the $Q\approx 1.1$~\AA$^{-1}$ wave-vector with a very steep dispersion at such low energy transfers.  Figure.~\ref{IRIS-HB3-figure}(b) shows the difference of the elastic scattering intensity between the disordered and ordered phases as measured using IRIS and HB3 illustrating the low-temperature powder magnetic Bragg peaks in Na$_3$RuO$_4$.  These data illustrate that the excitations are dispersing directly out of the magnetic Bragg peaks.  The data in Figs.~\ref{IRIS-HB3-figure}(a) show no indication of a gap in the magnetic spectrum down to $\sim$ 250~$\mu$eV.

\section{Discussion}

The high temperature magnetic susceptibility fitted using the finite cluster models for
T $>$ 30 K (where there is no long range order). Table \ref{Mag_Sus_Parameters}
shows the magnetic interactions determined from these fits for the three
models considered. We also compare the extracted magnetic interaction parameters
to the Curie-Weiss temperature. The Curie-Weiss temperature in a mean
field approximation is
\be
\Theta_{W} = \frac{S(S+1)}{3}J_0,
\ee
where, in this case, S = $\frac{3}{2}$ and $J_0$ is the sum of the magnetic
exchange constants \cite{Ash76}.  As shown in Fig. \ref{ChivsT}, all three cluster
models qualitatively reproduce the high temperature susceptibility data, and only
deviate strongly close to the transition temperature. However, a comparison of the
calculated Curie-Weiss temperatures based upon the double dimer model is
more consistent with the experimental Curie-Weiss temperature,
$|\Theta_{W}|=14.0$~meV (antiferromagnetic), illustrated in the inset of
Fig.~\ref{ChivsT}.  Because the Curie-Weiss temperature is proportional
to a sum of exchange constants, the presence of inter-cluster exchange
could significantly effect the estimate value of $\Theta_{W}$.  For example,
there are 28 bonds with distances of five to six angstroms between the Ru
sites in one cluster and the Ru sites in all neighboring clusters. If inter-cluster
interactions are large, the estimated Curie-Weiss temperature would deviate
significantly from the values quotes in Table~\ref{Mag_Sus_Parameters}.

\begin{figure}
\includegraphics[scale=0.95]{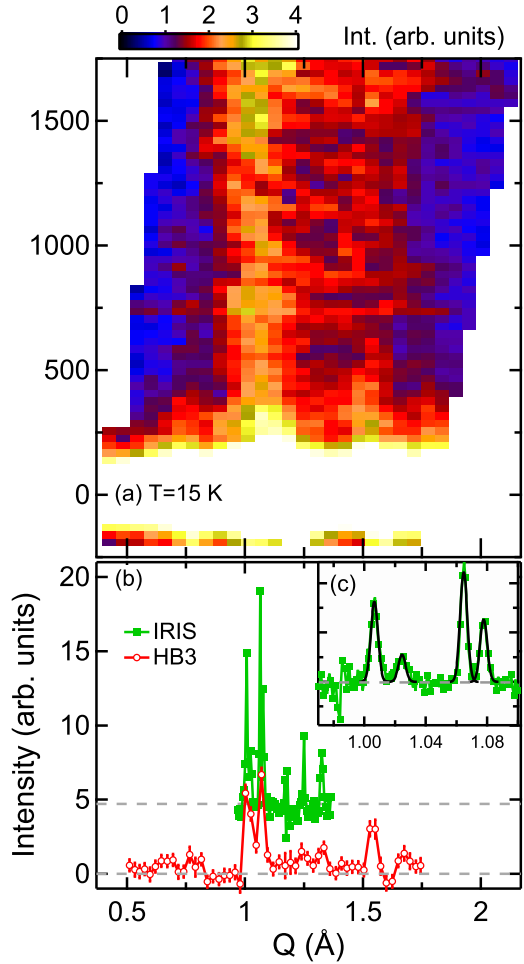}
\caption{(a) $T=15$~K Powder INS scattering intensity from Na$_3$RuO$_4$ as a function of energy and wave-vector transfer measured
using the IRIS spectrometer at ISIS.  (b) Intensity as a function of wave-vector transfer for data acquired at both IRIS and HB3. IRIS data were acquired at $T=8$~and~$30$~K using the high resolution diffraction banks.  HB3 data were acquired in same configuration as Fig.~\ref{braggcontour} at $T=8$~and~$30$~K. Both data sets indicate clear peaks at 1.0 and 1.07 \AA$^{-1}$.  Panel (c) shows the high resolution IRIS data fitted to a series of single width Gaussian peaks at $Q=1.006(1)$, 1.025(4), 1.065(1) and 1.078(1) \AA$^{-1}$.}
\label{IRIS-HB3-figure}
\end{figure}

The inelastic neutron scattering energies and intensities can also be calculated
using the exchange values determined from the magnetic susceptibility
(Table~\ref{Mag_Sus_Parameters}) and the ion positions given by Regan
$et~ al.$\cite{regan2005} The three models predict INS observable energy
gaps of 2.96, 3.08, and 12.16 meV for the lozenge model, 1.33, 3.49, and
23.53 meV for the coupled dimer model, and 2.47 and 9.16 meV for the
double dimer model.  We have already noted that the data shows magnetic
excitations at approximately 5.0 and 9.8 meV, \textit{cf.} Fig.~\ref{IvsE}.
Figures~\ref{MARITemp} and ~\ref{IRIS-HB3-figure} also show evidence for a spin-wave emerging
from the antiferromagnetic Bragg peak in the vicinity of the  $Q~\approx~1~$\AA$^{-1}$
for $T<T_N$.  Although a clear transition to long-range magnetic order is evident at low temperatures, it is nonetheless reasonable to examine the properties of these excitations as observed in INS especially their wave-vector dependence,
since these are characteristic indicators of the nature of the interactions.

\begin{figure}
\includegraphics[width=3.75in]{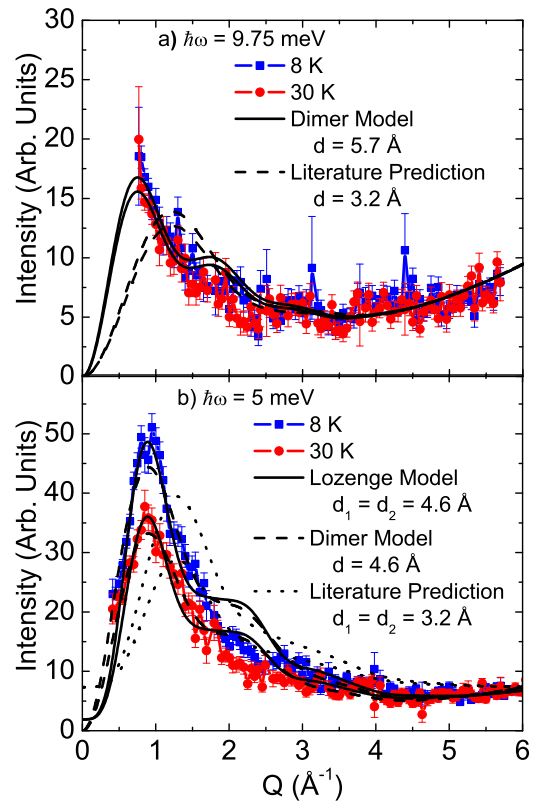}
\caption{a) INS intensity versus momentum transfer of the 9.8 meV
excitation in Na$_3$RuO$_4$ at T = 8 K(blue squares) and 30 K
(red circles); the black solid lines are fits to the dimer model and
the gray dashed lines are predictions given fixed physical distances.
b)INS intensity versus momentum transfer at T = 8 (blue squares) and
30 K (red circles) for the 5.0 meV excitation. The solid black line is the
lozenge model fit, the dashed gray line is the dimer model fit, and dotted
light gray line is a prediction using physical distances (The data were
taken using the MARI spectrometer at ISIS, as described in the text).}
\label{IvsQ}
\end{figure}

On comparing the predicted excitation spectrum, using the magnetic susceptibility
data to determine the interaction strengths, the double dimer
model seems to give the most realistic description of the excitation energies.  We note
however
that the inter-cluster interactions which produce long-range order may significantly
affect the energy levels.  All three models predict an INS visible excitations below
3 meV and above 9 meV.  We also examine the wave-vector dependence of these
three models;  Figure \ref{IvsQ} shows constant energy scans performed above
and below $T_N$ at 5.0 and 9.8 meV energy transfer.  For comparison, we first
calculate the wave-vector dependence of the INS scattering intensity using
Eq.~\ref{eq:powderintensities} for the lozenge model geometry.  These
lineshapes (shown in Fig.~\ref{IvsQ}) are unable to account for the initial
rapid rise in scattering intensity at small wave-vectors, which implies that
there are significant exchange interactions between spins at larger separations
than are present in the lozenge model.  If the ionic distances are allowed to vary
freely in fitting the data, an interesting result emerges. For the 9.8 meV excitation,
the fitted dimer separation in a single dimer model is 5.66$\pm$0.10~\AA~
bond.
This separation agrees with the length of the $\gamma$-dimer.  This makes it unlikely
that the coupled dimer model is realistic, as it predicts a 23 meV $\gamma$-dimer
 excitation using the magnetic susceptibility parameter. The lozenge and double-dimer
 models with susceptibility-fitted parameters predict $\gamma$-dimer excitations at 12
 meV and 9 meV, respectively. The double dimer model is evidently closer to the
observed gap of 9.8 meV. Fitting the 5.0 meV data with a free dimer length in which,
both the lozenge and dimer models gives a length of 4.60$\pm$0.02~\AA, which does
not correspond to any Ru-Ru separation in the structure of Na$_3$RuO$_4$. We
conclude that the intercluster interactions are important enough to modify the energies
and wave-vector dependences of the excitations, so that the three simple dimer models
do not give a decent description of the excitations.

\begin{table}
\caption{\textbf{Exchange interactions estimated from the magnetic susceptibility and inelastic neutron scattering.} }
\begin{ruledtabular}
\begin{tabular}{c}
Curie-Weiss Fit \\
$|\Theta _w|$ = 14.0$\pm$0.2 meV \\
\end{tabular}
\begin{tabular}{l}
Magnetic Susceptibility
\end{tabular}
\begin{tabular}{lcccc}
 Model & $ J$ (meV) & $\alpha J$ (meV)& $\gamma J$ (meV)  & $|\Theta _w|$ (meV) \\
\hline \hline
Lozenge & 3.19$\pm$0.01 & 3.11$\pm$0.03 &   & 19.8$\pm$0.1 \\
\hline
Coupled& 3.7$\pm$0.1 & 4.6$\pm$0.3 & -4.8$\pm$1.2  & 18.3$\pm$1.3\\
Dimer &&&&\\
\hline
Double &  & 2.69$\pm$0.01 & 9.75$\pm$0.05  & 15.6$\pm$0.1\\
Dimer &&&&\\
\end{tabular}
\begin{tabular}{l}
Inelastic Neutron Scattering$^a$
\end{tabular}
\begin{tabular}{lcccc}
 Model & $J$ (meV) & $\alpha J$ (meV)& $\gamma J$ (meV) & $|\Theta _w|$ (meV)  \\
\hline \hline
Lozenge & 5.03$\pm$0.08 &  & 3.5$\pm$0.1 & 29.5$\pm$0.1 \\
\hline
Double &  & 5.03$\pm$0.08 & 9.8$\pm$0.2  & 18.5$\pm$0.1\\
Dimer \\
\end{tabular}
\end{ruledtabular}

\footnotemark{There is insufficient information to determine the interaction strengths in the coupled dimer model.}
\label{Mag_Sus_Parameters}
\end{table}

In Table~\ref{Mag_Sus_Parameters}, gives estimated magnetic interaction parameters
for the various models using the observed neutron scattering excitations at
$\hbar\omega \approx 5.03$~and $9.8$~meV as input. Since no third magnetic
excitation was observed to 30 meV, the coupled dimer model cannot be
uniquely constrained. Although the lozenge model also predicts three excitations,
two of those excitations involve dimers. Therefore, we can determine both
exchange constants using the neutron scattering results. Using these
exchange interactions, the lozenge model gives an estimated Curie-Weiss temperature
of 29.5 meV, more than double the observed value from magnetic susceptibility.
The double dimer model gives a value of $|\Theta _w|=18.5$~meV; this is closer
to the measured value, although it is still 30\%~larger than the observed value
(from the magnetic susceptibility).  If the excitations are indeed well described
by the double dimer model, then they demonstrate that the 5.0 and 9.8 meV
modes correspond to excitations of the $\alpha$- and $\gamma$-dimers, with
the nature of the $\alpha$-dimer excitation being significantly modified
coupled-cluster effects and the onset of long-range magnetic order.

\section{Conclusions}

In summary, we have given analytical results for the energy eigenvalues
and eigenstates for coupled dimers with general spin $S$ ions.
We derive analytical closed-form results for magnetic susceptibility
and inelastic neutron scattering excitation functions and their
wave-vector dependences for several tetramer models, and
compare our results to experimental data on the $S=3/2$
tetramer spin lozenge candidate, Na$_3$RuO$_4$.

On considering the observed magnetic susceptibility and
inelastic neutron scattering data and comparing these results
to several tetramer models, we first find that the Na$_3$RuO$_4$
data is not consistent with the spin-lozenge model of Regan
$et~al.$\cite{regan2005}.  Although no isolated tetramer model
is able to describe all of the thermodynamic and spectroscopic
measurements simultaneously, a double dimer model, with
bond lengths of $4.60$~\AA~ and $5.66$~\AA, does provide
a description of some aspects of the observed thermodynamic
properties and the inelastic neutron scattering measurements
on the two observed magnetic excitations. However, only one
spatial distance corresponds to  length seen in the material
shown in Fig. \ref{NaRuOStructure}.

Measurements of the heat capacity and elastic neutron scattering
data show that there are two distinct magnetic phase transitions
in this material, at $T\approx 23$ and $28$~K (This is the first
evidence for two low-temperature phase transitions in this material).
Clearly, an understanding the nature of these long-range ordered
magnetic phases will provide useful additional information regarding
the nature of magnetic interactions in Na$_3$RuO$_4$. We anticipate
that neutron diffraction measurements on single crystal samples of
Na$_3$RuO$_4$ will be the most useful next step in the experimental
studies of this material.

\section{Acknowledgments}

We would like to acknowledge the Joint Institute for Neutron Sciences for funding and support. We thank S. Nagler for helpful discussions. The research at Oak Ridge National Laboratory was sponsored by the Scientific User Facilities Division, Office
of Basic Energy Sciences, U. S. Department of Energy.


\begin{thebibliography}{harald}

\bibitem{Heis26} W. Heisenberg, Z. Phys. \textbf{38}, 411 (1926)

\bibitem{Gatt06}
D. Gatteschi, R. Sessoli, and J. Villain {\it Molecular Nanomagnets}
(Oxford University Press, 2006).

\bibitem{Dag96} E. Dagotto and T. M. Rice, Science \textbf{271}, 618 (1996).

\bibitem{Kah93} O.Kahn,
{\it Molecular Magnetism} (VCH Publishers, New York, 1993).

\bibitem{Nie00}
M. A. Nielsen and I. L. Chuang,
{\it Quantum Computation and Quantum Information}
(Cambridge, 2000).

\bibitem{Bar99} A. L. Barra, A. Caneschi, A. Cornia,
F. Frabrizide de Biani, D. Gatteschi, C. Sangregorio, R. Sessoli, and L. Sorace,
J. Am. Chem. Soc. \textbf{121}, 5302 (1999).

\bibitem{sha80} I. S. Shaplygin and V. B. Lazarev, Russ. J. Inorg. Chem. \textbf{25}, 1837 (1980).

\bibitem{shi04a} M. Shikano, C. Delmas, J. Darriet, Inorg. Chem. \textbf{43}, 1214 (2004).

\bibitem{shi04b} M. Shikano, R. K. Kremer, M. Ahrens, H.-J Koo, M.-H Whangbo, and J. Darriet, Inorg. Chem. \textbf{43}, 5 (2004).

\bibitem{regan2005}
K. A. Regan, Q. Huang and R. J. Cava, J. Solid State Chem. {\bf 178}, 2104 (2005).

\bibitem{cal66} A. Callaghan, C. W. Moeller, and R. Ward, Inorg. Chem \textbf{5}, 1572 (1966).

\bibitem{Rij99} J. T. Rijssenbeek, R. Jin, Y. Zadorozhny, Y. Liu, B. Batlogg, and R. J. Cava, Phys. Rev. B \textbf{59}, 4561 (1999).

\bibitem{Mae94} Y. Maeno, H. Hashimoto, K, Yoshida, S. Hishizaki, T. Fujita, J.G. Bernorz, and F. Lichtenberg, Nature \textbf{372}, 532 (1994).

\bibitem{cav04} R. J. Cava, J. Chem. Soc. Dalton Trans. \textbf{19}, 2979 (2004).

\bibitem{Reg06} K. A. Regan, Q. Huang, M. Lee, A.P. Ramirez, and R.J. Cava,
J. Solid State Chem. \textbf{179}, 195 (2006).

\bibitem{Alex03} A. Alexander, P. D. Battle, J. C. Burley, Daniel J. Gallon, Clare P. Grey, S. H. Kim, J. Mater. Chem.  \textbf{10}, 2612 (2003).

\bibitem{Mog04} K. M. Mogare, K. Friese, W. Klein, and M. Jansen,
Z. Anorg. Allg. Chem. \textbf{630}, 547 (2004).

\bibitem{Darr74}
J. Darriet and J. Galy, Bull. Soc. fr. Mineral. Cristallogr. \textbf{97}, 3 (1974).

\bibitem{regannote}
We believe there are two typographical errors in the
table of atomic coordinates in Ref.~\onlinecite{regan2005}.  The
$z$ fractional coordinate of site Na$_3$ and the
$y$ fractional coordinate of site O$_3$ should each be approximately
$0.5$.

\bibitem{Dril77} M. Drillon, J. Darriet, and R. Georges, J. Phys. Chem. Solids \textbf{38}, 411 (1977).

\bibitem{gib80} T. C. Gibb, R. Greatrex, N. N. Greenwood, J. Solid State Chem. \textbf{31}, 153 (1980).

\bibitem{Kam50} K. Kambe, J. Phys. Soc. Jpn. \textbf{5}, 48 (1950)

\bibitem{Squ78} G.L.Squires,
{\it Introduction to the Theory of Thermal Neutron Scattering}
(Dover, 1996).

\bibitem{Har05} J.T. Haraldsen, T. Barnes, and J. L. Musfeldt, Phys. Rev. B \textbf{71}, 064403 (2005).

\bibitem{Park03} N. G. Parkinson, P. D. Hatton, J. A. K. Howard, C. Ritter, F. Z. Chiend and M.-K Wue, J. Mater. Chem \textbf{13}, 1468 (2003).

\bibitem{Park05} N. G. Parkinson, P. D. Hatton, J. A. K. Howard, S. R. Giblin, I. Terry, C. Ritter, B.-H. Mokd and M.-K. Wud, J. Mater. Chem \textbf{15}, 1375 (2005).
    
\bibitem{mari}
M. Arai, A.D. Taylor, S.M. Bennington, and Z.A. Bowden,
in  \textit{Recent Developments in the Physics of Fluids},
W.S. Howells \& A.K. Soper eds, p. F291 (Adam Hilger, Bristol, 1992).

\bibitem{iris}
C.J. Carlile and M.A. Adams, Physica B \textbf{182}, 431 (1992).

\bibitem{Ash76} N. W. Ashcroft and N. D. Mermin,
{\it Solid State Physics}
(Thomson Learning, 1976).


\end{thebibliography}
\end{document}